\documentclass{article}
\renewcommand{\baselinestretch}{2}
\input epsf
\begin{document}

\renewcommand{\baselinestretch}{1}

\title{On matter-antimatter separation \\
 in open relativistic material system}
\author{Anatoli A. Vankov}
\date{Department of Physics, Astronomy, and Materials Science,\\
 Southwest Missouri State University, 901 South National Avenue,\\
 Springfield, Missouri, 65804, e-mail anv175f@smsu.edu}

\maketitle
\renewcommand{\baselinestretch}{2}
    
\begin{abstract}
 An open (having no physical boundaries) baryon symmetric system is 
 considered in a flat space-time. We assume that a space is uniformly
 filled with electromagnetic radiation and material objects, and the
 system is isotropic in any inertial reference frame. It means that a
 coordinate-momentum distribution of radiation and material objects
 does not depend on a reference frame, and the system should be in
 a state of chaotic relativistic motion. The dominant interaction
 processes are like-matter merge, unlike-matter annihilation and
 pair production. In this approximation we came to the conclusion
 that matter and antimatter exist in a form of mix of material
 objects characterized by a stationary baryon symmetric mass
 distribution in a broad mass range. In other words, a
 matter-antimatter space separation takes place.
\end{abstract}

\section{Formulation of the problem}

 A matter-antimatter mix is generally expected to annihilate. However,
 under certain conditions a trend might appear, which leads to a
 matter-antimatter space separation. The problem was discussed
 elsewhere (see, for example \cite{Omnes,Klein,Stecker,Fichtel}).
 In the present work an original approach is discussed. Our
 formulation of the problem is as follows.

\begin{itemize}
\item[a.]
 An {\em open system} of material objects and radiation
 is considered in a flat space-time. There is no phisical cause for
 preference in choosing inertial reference frames. Thus, the system
 is assumed to be baryon symmetric, uniform and isotropic on average
 {\em in any reference frame}. It leads to specific requirement
 for a matter coordinate-momentum distribution, as shown further.
\item[b.]
 Material objects are characterized by a mass distribution (in a
 unit volume) $N(m)$, which is the same for matter and antimatter.
 Baryon charge is conserved. In addition, the system is characterized
 by a uniform coordinate-momentum distribution (in a unit volume).
\begin{equation}
 f({\rm x},{\rm p})= Cons\, d{\rm x}\,d{\rm p} ~,
\end{equation}
 where ${\rm x}$, ${\rm p}$\hskip.1in --- space and momentum
 3-vectors.
\item[c.]
 All known physical interactions are allowed between system constituents.
\end{itemize}

 The above formulation of the problem reflects the idea of a generalized
 matter transport equation.  In the present work a treatment of the
 problem will be as much simplified as possible to concentrate our
 attention on the question of matter-antimatter separation. We do
 not specify the physical nature of material objects.  Any concrete
 astrophysical picture of a matter structure might be embedded, if
 needed, into the frame under consideration. In general, an object
 is meant to be a free microscopic or macroscopic particle, free solid
 body or a gravitationally linked system (particle cloud, multi-body
 association, galaxy etc). We assume that a free object is made of
 either matter or antimatter. Hence, an object is characterized by
 its mass at rest and baryonic type, its internal evolution being
 ignored. For the purpose of this work it seems sufficient to consider
 object-object interaction in general terms of a random collision
 followed by a formation of a compound system, which may disintegrate
 into new objects with channel probabilities of fragmentation,
 annihilation, and merge processes. We assume that matter-antimatter
 annihilation results in gamma radiation being a source of a following
 pair production. It means that annihilated matter is regenerated in
 the ``first group'' of a mass distribution. The ``first group'' may be
 referred, in principle, to elementary particles as  a ``seed matter''
 and plays a role of a source term in a kinetic equation.

 Next comments concern relativistic properties of an open system. 
 First of all, we need to explain the term ``relativistic material
 system''. A uniform isotropic matter distribution in a flat space
 is characterized by a uniform coordinate-momentum distribution
 function (1). It is Lorentz-invariant, that is unchanged if measured
 in any inertial reference frame. Such a state of matter can be treated
 as a ``maximal chaos'' in terms of Bayesian approach for description
 of relativistic gas \cite{Synge}. Actually, this is an approximation
 of non-interacting particles. It is worth noting that the relativistic
 gas model has nothing in common with Friedmann-Lemaitre expanding
 universe model \cite{Friedmann,Robertson,Peebles}, which treats matter
 like dust-like matter-made particles with small relative velocities in
 an observer's vicinity. The expanding universe itself plays a role
 of the absolute reference frame. In our approach an opportunity of
 introducing the absolute reference frame is denied. However, the
 above approximation of non-interacting particles leads to an infinite
 energy density. In a more rigorous relativistic gas model one has
 to take into account that physical processes locally observed under
 conditions of an open system should be characterized by retarded
 casual connections with the rest of the space, and inertial systems
 must be referred to a limited space-time volume depending on how
 precise they are needed to be defined. Hence, a ``realistic''
 coordinate-momentum distribution must be characterized by some
 space-energy correlation and must have a smooth cut-off at however
 high energy range, or an effective temperature parameter. Now any
 local energy density will be found limited and invariant in a broad
 set of inertial systems (the higher temperature, the broader set).
 We assume that the temperature is well above the threshold needed
 for pair production process being effective.

 Under above conditions annihilation and pair production processes
 have to be  balanced, for there is no energy dissipation in the
 open system. In other words, the system is expected to be in a
 self-sustained baryon symmetric state. If a mass distribution
 function is not degenerated into the ``first source group'' one
 can state that a matter-antimatter space separation takes place.
 The mass distribution function should be found from a material
 balance (kinetic) equation.

\section{Model kinetic equation}

 Let a mass distribution function for like matter objects (in a
 unit volume) be given in a mass group form for matter  $m_i$ or
 antimatter $m_i^*$:
\begin{equation}
 N_i = N(m_i)  {\hskip.2in}or{\hskip.2in} N_i^* = N(m_i^*)
\end{equation}
 with a total mass conserved:
\begin{equation}
 M = \sum m_iN_i{\hskip.2in} M^* = \sum m_i^*N_i^*{\hskip.2in} M = M^* ~.
\end{equation}

 Next we use so called {\em one velocity approximation}, that is an
 equation being averaged over momentum distribution. Let us introduce
 a generation rate $G_i$ of matter-made objects in a group $i$ in a
 result of matter-matter $j-k$ type object collisions. If widths of
 groups are narrow enough we may ignore interactions inside groups.
 \begin{equation}
 G_i = \sum_j\sum_k<N_jN_kv\sigma(m_j,m_k)K([m_j,m_k]\rightarrow m_i)> ~,
 \end{equation}   
 here the brackets $<>$ symbolize averaging over momentum distribution,
 $v$ is a relative velocity, $\sigma$ is an object-object collision
 cross-section, and $K([m_j,m_k]\rightarrow m_i)$ is a ``channel function'',
 describing a probability of an object to appear in a group $i$ in a
 result of decay of compound system [$m_j$,$m_k$]. Masses are conserved
 due to a proper K-function normalization.

 Similarly, an object generation rate $G_i^*$ resulting in
 matter-antimatter type collision (* symbolize ``antimatter
 participant'', as before) may be given:
\begin{equation}
 G_i^* = \sum_j\sum_{k^*}<N_jN_k^*v\sigma(m_j,m_k^*)
 K([m_j,m_k^*]\rightarrow m_i)> ~.
\end{equation}

 In this case, the K-function is normalized to ensure a correct mass
 balance, annihilated mass in a form of gamma-radiation being taken
 into account.

 Choosing a proper summation rule we come to expressions for a removal
 rate $R_i$ and $R_i^*$ in matter-matter and matter-antimatter types of
 collision, correspondingly.
\begin{equation}
 R_i = \sum_j\sum_k<N_iN_jv\sigma(m_i,m_j)K([m_i,m_j]\rightarrow m_k)>
\end{equation} 
\begin{equation}
 R_i^* = \sum_j^*\sum_{k}<N_iN_j^*v\sigma (m_i,m_j^*)
 K([m_i,m_j^*]\rightarrow m_k)>
\end{equation}

 As is said above, an annihilation rate $G_\gamma$ is followed by a
 matter recreation in the first group, $G_1(m_r)$, masses of radiation
 $m_\gamma$ and recreated matter $m_r$ being in balance:
\begin{equation}
 G_1(m_r)=G_\gamma = \sum_j\sum_{k^*}<N_jN_k^*v\sigma(m_j,m_k^*)
 K([m_j,m_{k^*}]\rightarrow m_\gamma) > ~.
\end{equation}

 Finally, we have a stationary kinetic equation:
\begin{equation}
 G_i+G_i^*+R_i+R_i^* = \delta_{1i} G_1(m_r)  \hskip.4in {(i=1,2,3,...)} ~,
\end{equation}
 where $\delta_{1i}=1$ for $i=1$, and $\delta_{1i}=0$ for $i\not=1$.

 Due to a symmetry of physical properties of matter and antimatter one
 can get the same equation starting considering generation/removal rates
 for antimatter-made objects. The equation (9) may be written in an
 integral form. It is easy to see that the equation has a non-trivial
 solution if a merge channel providing a transfer from a mass group $i$
 to a mass group $i+1$ is open. Since microscopic and macroscopic merging
 processes are known from conventional Physics we may conclude that the
 above equation has a physical meaning.

\section{Discussion}

 To demonstrate a solution we realized the Monte Carlo method for
 K-functions reduced to delta-functions. In other words, the
 fragmentation process has been ignored. We checked that it does
 not appreciably influence the form of solution though a computing
 time significantly increases. Obviously, a solution depends on a
 collision cross-section as a function of an object mass. Four
 variants were studied: constant cross-section, cross-section,
 proportional to the mass of a target object, cross-section proportional
 to both the mass of target and the mass of incident object, finally
 cross-section proportional to the squared mass of the target object.
 The last two variants were expected to have only slightly different
 solutions. The variants seem to include typical types of mass-dependent
 cross-sections of body-body physical interactions. However, concrete
 energy-dependent cross-sections for specific interactions may differ
 significantly in magnitude in different mass intervals. We do not know
 to what extent the solutions are influenced by ``one velocity
 approximation''. This important question needs a special study on
 open relativistic matter system properties, their dependence on energy
 density, in particular. Besides, taking into account an object internal
 evolution might be important also if the results were used for their
 speculative applications concerning astrophysical problems \cite{Vankov}.
 All these questions are out of the scope of the present work.

 The main conclusion we would like to emphasize is that an open
 steady-state isotropic system consisted of baryon symmetric uniformly 
 distributed matter should be characterized by a relativistic momentum
 distribution with an effective high temperature parameter. Subsequently,
 a space matter-antimatter separation takes place resulted from random
 interactions within the system. The possible scale of separation may
 be seen from the statistical assessments of stationary mass distribution
 functions in Fig.~1. The picture gives us the idea of a solution of a
 simple model kinetic equation describing the above system: the solution
 may be approximated by a function $1/m^n$ with $n$ ranging between 2
 and 3.5, depending on model parameters. Apparently, there might be a
 trend of decreasing $n$ for asymptotic solution. The latter should be
 found by analytical method. The result shows that an open baryon
 symmetric matter system reveals a structure containing matter or
 antimatter-made objects of any size. There is no material limit for
 their construction because of the system being unlimited in space.
 The bigger size (age) objects have, the less their population is.
 Further computer simulations with more sophisticated models might
 show more physically meaningful features of the baryon symmetric
 open system.

\begin{figure}
\epsfbox{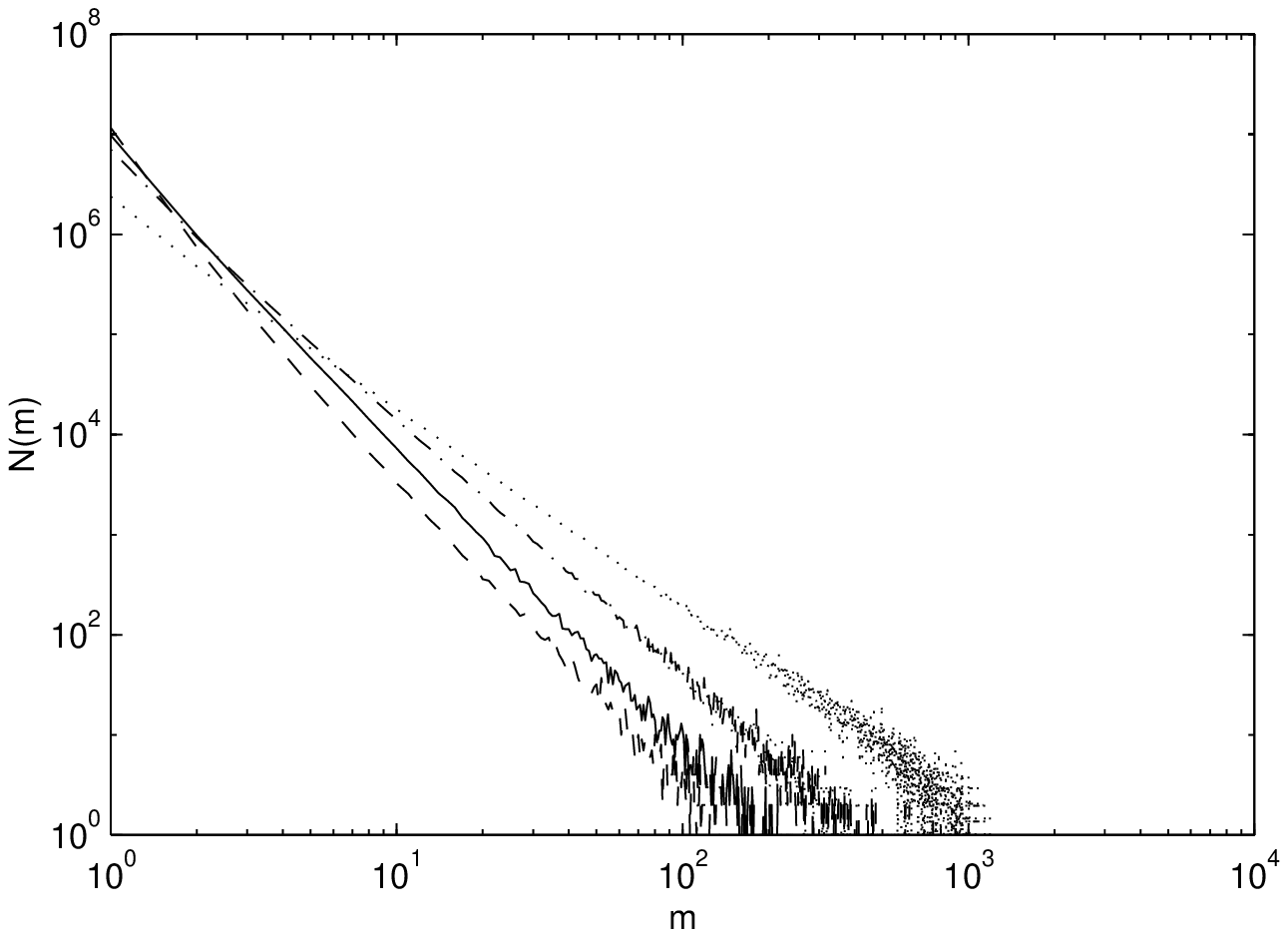}
\caption{\leftline{Mass distribution after 250 million random tests for
 variants~:}
 \leftline{(~$\cdots$~) interaction cross-section is constant,}
 \leftline{(-- $\cdot$ --) proportional to the mass of a target object,}
 \leftline{(~~---~~) proportional to the mass of incident object and to
  the one of target object,}
 \leftline{(-- -- --) proportional to the mass squared.}}
\end{figure}

\end{document}